# Using Spatial Correlation in Semi-Supervised Hyperspectral Unmixing under Polynomial Post-nonlinear Mixing Model

Fahime Amiri[a], Mohammad Hossein. Kahaei [b*]

*[a,b] School of Electrical Engineering, Iran University of Science and Technology, Narmak, Tehran, Iran*

[*] Corresponding author, e-mail address: kahaei@iust.ac.ir

Abstract: This paper presents a semi-supervised hyperspectral unmixing solution that integrate the spatial information in the abundance estimation procedure. The proposed method is applied on a nonlinear model based on polynomial postnonlinear mixing model where characterizes each pixel reflections composed of nonlinear function of pure spectral signatures added by noise. We partitioned the image to classes where contains similar materials so share the same abundance vector. The spatial correlation between pixels belonging to each class is modelled by Markov Random Field. A Bayesian framework is proposed to estimate the classes and corresponding abundance vectors alternatively. We proposed sparse Dirichlet prior for abundance vector that made it possible to use this algorithm in semi-supervised scenario where the exact involved materials are unknown. In this approach, we just need to have a large library of pure spectral signatures including the desired materials. An MCMC algorithm is used to estimate the abundance vector based on generated samples. The result of implementation on simulated data shows the prominence of proposed approach.

## Introduction

Hyperspectral images are mentioned as images have been taken in a hundred of spectral bands in remote sensing area [Keshava 2002]. They have been used in various applications including agricultural and environmental monitoring [Somers 2009], mineral exploration [Settle 1993], military surveillance [Chang 2000], and so on [Liu 2014]. Due to physical limitations of imaging devices, each hyperspectral pixel is mixture of reflectance signatures of materials in the field of view, named endmembers [Vane 1993]. Spectral unmixing denotes as decomposition of a mixed pixel into a group of pure spectral signatures and their corresponding proportions [Keshava 2002]. Most of unmixing algorithms assume Linear Mixing Model (LMM) where a hyperspectral pixel is indicated by a convex mixture of some endmembers [Heinz 2001; Eches 2010; Fu 2016; Akhtar 2017].

Although the LMM has comprehensive usage in unmixing application, it has drawbacks in images of special materials [Keshava 2002]. In this case, the nonlinear mixing models are introduced to solve the linear problems. It has been shown that the nonlinear unmixing approaches have better result than linear ones [Yu 2016]. Most researches consider bilinear mixing effect to simulate the nonlinear mixing in hyperspectral unmixing [Imbiriba 2016; Halimi 2011; Somers 2009]. The bilinear model assumes that light beams go through multiple reflections. In [Dobigeon 2008], an iterative technique for estimating the endmember matrix has been proposed under assumptions that linear mixture of endmembers are present within at least a small part of the image. Above algorithms need to know the number of endmembers. Generalized Bilinear Mixing model (GBM) is one of the most common bilinear models [Halimi 2011]. Along bilinear mixing model, other nonlinearities like intimate mixture [Hapke 1981], kernel-based [Liu 2009] and neural network based approaches [Altmann 2011] are considered too. Another interesting nonlinear mixing models is polynomial postnonlinear mixing model (PPNMM) [Altmann 2012] were was originally proposed in source separation problem [Babaie-Zadeh 2001] to solve supervised unmixing. An unsupervised version of [Altmann 2012] was also represented in [Altmann 2014], however a third-party nonlinear endmember extraction algorithm (EEA) [Heylen 2011] was needed to extract the mean vector of endmembers. Unfortunately, most nonlinear unmixing approaches are supervised; i.e., the exact endmembers presented in the image are assumed to be known. However, the unsupervised approaches relied on the EEAs that most of them are based on the LMM [Winter 1999; Nascimento 2005; Chaudhry 2006] and thus may result incorrect endmembers in case of nonlinear mixtures.

While a nonlinear EEA has been proposed in [Heylen 2011] to extract endmembers from the data, this algorithm could not lead to accurate spectral signature under absence of pure pixels in the image (as most linear EEAs), and the effectiveness of using manifold learning methods on real data has not been confirmed yet. Although [Dobigeon 2008] has been used to estimate the number of endmembers, it relied on Principal Component Analysis (PCA) where assumes linearity inherently.

Above unmixing algorithms where applied on pixel by pixel context independently, do not take advantage of the spatial correlations between the neighbour pixels of the hyperspectral image. Spatial information are valuable data in improving unmixing and classification accuracy [Plaza 2002; Rogge 2007; Zhao 2017]. In Bayesian estimation structures, Markov Random Fields (MRFs) are widely used for modelling spatial correlation in images [Eches 2013; Tarabalka 2010; HongLei 2013]. In unmixing problem MRFs have also been used to improve the unmixing accuracy [Eches 2011, Chen 2017]. These approaches were based on linear mixing assumption.

In this paper, we propose the Sparse Dirichlet Prior with PPNMM (SDP-PPNMM) algorithm in a semi-supervised manner that means we do not need any EEA and the lack of knowledge about pure endmembers is compensated just by selecting suitable priors. Moreover, the proposed algorithm doesn't require the presence of pure pixels in the observed image. We assume that a large library of endmembers is available, which is a realistic assumption due to collecting a wide variety of spectral signature of various common materials during few decades. The U.S. Geological Survey (USGS) Library is one of these publicly available libraries which has been taken over 22 years covers more than 1,300 spectral signature of so many materials [Clark 2007]. Since, in each hyperspectral pixel only a small number of endmembers against the extremely large library are contributed, the abundance vector could be sparse. In this paper a sparse Dirichlet prior is proposed for abundance vector. Thus, unmixing and endmember selection from a large dictionary are executed simultaneously. We also make use of MRF to profit from spatial correlation in powerful unmixing algorithm Bayesian PPNMM and thus major improvement in unmixing accuracy is achieved. Using MRF during unmixing procedure, made it possible to classify the hyperspectral image to known number of classes wherein the abundance vector and the nonlinearity term is the same in each class. Thus our algorithm performs not only endmember selection during unmixing process in a flexible generalization of LMM, PPNMM, but also make classification of the hyperspectral pixels.

**Problem Formulation**

It is important to note that due to contribution of only a small number of endmembers of an extremely large library in each hyperspectral pixel, the abundance vector could be sparse. Accordingly, in this paper we consider this case for estimating abundance vectors in a Bayesian sense. In this way, unmixing and endmember selection from a large library are performed simultaneously.

To elaborate, in a nonlinear mixing model, a hyperspectral pixel is defined as a nonlinear function of a linear mixture of endmember signatures affected by noise term as

$$\mathbf{y} = g(\sum_{r=1}^{R} a_r \mathbf{m}_r) + \mathbf{n} = g(\mathbf{Ma}) + \mathbf{n} \quad (1)$$

where $\mathbf{y}$ is an $L$-dimensional hyperspectral pixel, $\mathbf{m}_r$ is the spectral signature of the $r$th endmember in the library $\mathbf{M}$, $a_r$ is the corresponding abundance, $R$ is the number of endmembers in the library, $\boldsymbol{g}$ is a nonlinear transformation, and $\mathbf{n}$ is additive white Gaussian noise with zero-mean and variance $\sigma^2$:

$$\mathbf{n} \sim \mathcal{N}(\mathbf{0}, \sigma^2 \mathbf{I}_L) \qquad (2)$$

where $\mathbf{I}_L$ denotes an $L \times L$ identity matrix.

Here, a second-order polynomial is employed for the nonlinear function $\boldsymbol{g}_b$. It has been shown that second order polynomial is an appropriate approximation for nonlinear mixing models [Nascimento 2009], since higher order terms are negligible and could be merged in noise term. A PPNMM [Altmann 2012], is given by

$$\boldsymbol{g}_b: \mathbb{R}^R \times \mathbb{R} \to \mathbb{R}^L$$
$$(\mathbf{a}, b) \to \mathbf{Ma} + b(\mathbf{Ma}) \odot (\mathbf{Ma}) \qquad (3)$$

where $\odot$ is the Hadamard product. This model contains both bilinear and linear models. If $b = 0$, the model is simplified to LMM and thus the result would be at least as good as linear ones. According to [Altmann 2012], using $b$ as a single amplitude parameter for the nonlinear term, lower complexity is achieved.

Abundances has some consideration in practise. There are two constraints known as non-negativity and sum-to-one that mean each element of abundance vector should not be smaller than zero and sum of the abundance fractions should be equal to one:

$$a_r \geq 0, \ r = 1, \dots, R, \sum_{r=1}^{R} a_r = 1; \qquad (4)$$

**Spatial Correlation Formulation**

Previous efforts exhibited that MRFs are interesting tools for modelling spatial correlation used in hyperspectral image classification and segmentation in Bayesian framework [Chen 2017]. We utilize Potts-Markov random field [Eches 2011] to contribute spatial correlation with first order neighborhood pixels. Such a neighborhood, considers near vertical and horizontal neighbors. To define MRF, the hyperspectral images should classify to particular classes where pixels belonging to in are similar in the sense of abundance vector.

Suppose that the variables $\mathbf{c} = [c_1, \ldots, c_P]$ shows pixel classes for pixels $1, \ldots, P$ respectively, where defined on the set $\{1, \ldots, K\}$ and $K$ is number of classes. The set of variables $\{c_1, \ldots, c_P\}$ indicate a random field. MRF is a kind of random fields where the conditional distribution of $c_i$ given other pixel labels $\mathbf{c}_{\setminus i}$ only depends on its neighbor ones $\mathbf{c}_{\nu(i)}$:

$$f(c_i|\mathbf{c}_{\setminus i}) = f(c_i|\mathbf{c}_{\nu(i)}) \quad (5)$$

where $\nu(i)$ is neighborhood of pixel $i$, and $\mathbf{c}_{\setminus i} = \{c_j, j \neq i\}$. Using Potts-Markov model, based on Hammersly-Clifford theorem which relates the MRFs to Gibbs distribution, the pdf of random field $\mathbf{c}$ is written as follows:

$$f(\mathbf{c}) = \frac{1}{G(\beta)} \exp\left(\left[\sum_{p=1}^{P} \sum_{p' \in \nu(p)} \beta \delta(c_p - c_{p'})\right]\right) \quad (6)$$

where $\beta > 0$ is known as granularity coefficient, $G(\beta)$ is normalizing constant or partition function, and $\delta(.)$ Is the Kronecker function where $\delta(x) = 1$ if $x = 0$, and $\delta(x) = 0$ otherwise.

The Markov Chain Monte Carlo (MCMC) sampler used to generate samples based on $f(\mathbf{c})$. As you know, the Gibbs sampler does not depend on constant coefficient [Eches 2011], so the normalizing constant can be ignored. The parameter $\beta$ indicates how much a class is homogenous. The large $\beta$ leads to more homogenous class map and the very small $\beta$ tends toward a near noisy class map.

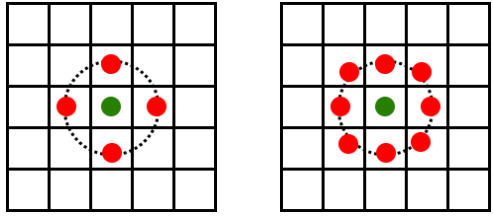

**Figure 1** 4-pixel and 8-pixel neighborhood

In this paper we consider 4-pixel first order neighborhood structure, so according to [Eches 2011], a fixed moderate value of $\beta$ is enough to avoid from trapping in a local optimum. For larger neighborhood structure, smaller value of $\beta$ is appropriate.

**Hierarchical Bayesian Framework**

In this section the likelihood function of observed pixel is computed based on PPNMM for the hyperspectral unmixing and the priors are considered to calculate the posteriors of unknown parameters. We utilize the hierarchical Bayesian solution. Our motivation is to select a proper prior for the abundance vector which leads to not using any EEA. Also we benefit the spatial correlation in nonlinear mixing model to enhance the unmixing performance.

Assuming Gaussian noise, the likelihood function of the mixed pixel is normally distributed denoted as:

$$f(\mathbf{y}_p | c_p = k, \mathbf{a}_k, b, \sigma^2) = \left(\frac{1}{2\pi\sigma^2}\right)^{\frac{L}{2}} \exp\left(-\frac{\|\mathbf{y}_p - g_b(\mathbf{M}\mathbf{a}_k)\|^2}{2\sigma^2}\right) \qquad (7)$$

where $\mathbf{y}_p$ is the observed hyperspectral pixel $p$ and $c_p$ is its corresponding class label. Unknown parameters $\mathbf{a}, b,$ and $\sigma^2$ should be estimated. Due to independency between noise elements $\mathbf{n}_p$ for pixels $p = 1, \ldots, P$ we have:

$$f(\mathbf{Y}|\mathbf{c}, \mathbf{A}, b, \sigma^2) = \prod_{p=1}^{P} f\left(\mathbf{y}_p | c_p, \mathbf{a}_{c_p}, b, \sigma^2\right) \qquad (8)$$

*Prior Selection*

In this section the prior distributions that have been chosen for unknown parameters $\boldsymbol{\Psi} = \{\mathbf{c}, \mathbf{A}, b, \sigma^2\}$ are described. In addition, a hyperparameter could be introduced and a proper prior would be assigned to it.

*Pixel Class Prior:*

As explained earlier, the prior pdf of the pixel class vector is a Potts-Markov random field with first-order neighborhood as [Eches 2011]. The parameter $\beta$ for the distribution is set to 1.1.

*Abundance Prior:*

Due to the two physical constraints, sum-to-one and non-negativity of the abundance vector, we propose the sparse symmetric Dirichlet distribution [Ng 2011] as a prior for the abundance vector $\mathbf{a}_k$ of class $k$ as

$$f(\mathbf{a}_k) \sim \mathcal{D}(\mathbf{a}_k; \eta) = \frac{\Gamma(\eta R)}{\Gamma(\eta)^R} \prod_{r=1}^{R} a_{k,r}^{\eta-1} \qquad (9)$$

where $\Gamma(.)$ is the Gamma function and $\eta$ exhibits the concentration parameter. The joint pdf of abundance vector of all $K$ class is equal to $f(\mathbf{A}) = \prod_{k=1}^{K} f(\mathbf{a}_k)$. This distribution presents a sparse behavior for $\eta < 1$ and corresponds to a uniform distribution over the standard $(R-1)$-simplex for $\eta = 1$. The latter case is commonly used in unmixing problems [Altmann 2012]. Note that this case is limited to supervised unmixing applications in which the exact endmembers must be known. In more realistic scenarios, however, exact mixing endmembers are unknown and only a large spectral library is given and the concentration parameter plays an essential role. Here, accordingly we consider $\eta < 1$ and show that this case is suitable for a wide range of applications.

*Noise Variance Prior*

We use a non-informative Jeffreys' prior for noise variance $\sigma^2$ [Altmann 2012] as follows:

$$f(\sigma^2) \propto \frac{1}{\sigma^2} \mathbf{I}_{\mathbb{R}^+}(\sigma^2) \qquad (10)$$

where $\mathbf{I}_{\mathbb{R}^+}(.)$ is the indicator function defined on the positive real values:

$$\mathbf{I}_{\mathbb{R}^+}(x) = \begin{cases} 1, & \text{if } x \in \mathbb{R}^+ \\ 0, & \text{otherwise.} \end{cases} \qquad (11)$$

*Nonlinearity Coefficient Prior*

For the unknown parameter $b$, the following priors is assigned:

$$b|\sigma_b^2 \sim \mathcal{N}(0, \sigma_b^2) \qquad (12)$$

where $\sigma_b^2$ is a hyperparameter for which the Inverse-Gamma prior is

$$\sigma_b^2 \sim \mathcal{IG}(\gamma, \rho) \qquad (13)$$

where $(\gamma, \rho)$ are set to $(1, 0.01)$ according to [Altmann 2012].

*Derivation of Posterior Distribution*

Using the likelihood of mixed pixel and assumed priors, the joint posterior distribution of all unknown variables $\boldsymbol{\theta} = \{\boldsymbol{\Psi}, \sigma_b^2\}$ would be extracted by hierarchical structure as:

$$f(\boldsymbol{\theta}|\mathbf{Y}) \propto f(\mathbf{Y}|\boldsymbol{\Psi}) f(\boldsymbol{\Psi}|\sigma_b^2) f(\sigma_b^2) \qquad (14)$$

which means $f(\boldsymbol{\theta}|\mathbf{Y})$ is proportional to the product of the likelihood function by the priors and hyperpriors.

$f(\mathbf{\theta}|\mathbf{Y})$

$$\propto \frac{1}{\sigma^2}\left(\frac{1}{\sigma_b^2}\right)^{\frac{3}{2}+\gamma} f(\mathbf{Y}|\mathbf{c},\mathbf{A},b,\sigma^2) \times \exp\left(-\frac{b^2+2\rho}{2\sigma_b^2}\right) \times \mathbb{P}(\mathbf{c}) \times \prod_{k=1}^{K}\prod_{r=1}^{R} a_{k,r}^{\eta-1} \quad (15)$$

where

$$f(\mathbf{Y}|\mathbf{c},\mathbf{A},b,\sigma^2) = \prod_{p=1}^{P} f\left(\mathbf{y}_p|c_p,\mathbf{a}_{c_p},b,\sigma^2\right) \quad (16)$$

As seen in (15), high complexity of the derived posterior distribution make it impossible to obtain closed-form statement to derive the MMSE or MAP estimates for the unknown parameter $\mathbf{\theta}$. Under this condition, we make use of MCMC sampler to generate the samples distributed according to (16) and then to approximately apply Bayesian estimators [Robert 2004] to these samples.

*Metropolis-Within-Gibbs Sampler*

In this section we investigate the Metropolis-within-Gibbs sampler to generate samples iteratively according to distributions $f(\mathbf{\theta}|\mathbf{Y})$ wherein there exist unknown parameters $\{\mathbf{c},\mathbf{A},b,\sigma^2\}$ and one unknown hyper-parameter $\sigma_b^2$.

First, the conditional distributions $\mathbb{P}\left(c_p|\mathbf{\theta}_{\setminus c_p},\mathbf{Y},\mathbf{M}\right)$, $f\left(\mathbf{a}_{c_p}|\mathbf{\theta}_{\setminus \mathbf{a}_{c_p}},\mathbf{Y},\mathbf{M}\right)$, $f(b|\mathbf{\theta}_{\setminus b},\mathbf{Y},\mathbf{M})$, $f(\sigma^2|\mathbf{\theta}_{\setminus \sigma^2},\mathbf{Y},\mathbf{M})$ and $f\left(\sigma_b^2|\mathbf{\theta}_{\setminus \sigma_b^2},\mathbf{Y},\mathbf{M}\right)$ should be derived and then, samples be generated for $\mathbf{c},\mathbf{A},b,\sigma^2$ and $\sigma_b^2$, respectively.

*Conditional distribution of $\mathbf{a}_{c_p}$*

For each pixel $p$, the Bayes theorem yields

$$f\left(\mathbf{a}_{c_p}|\mathbf{\theta}_{\backslash \mathbf{a}_{c_p}}, \mathbf{Y}, \mathbf{M}\right) \propto \prod_{p \in \mathbf{p}_k} f\left(\mathbf{y}_p | c_p = k, \mathbf{a}_k, b, \sigma^2, \sigma_b^2\right) f(\mathbf{a}_k)$$

$$\propto \prod_{p \in \mathbf{p}_k} \exp\left[-\frac{1}{2\sigma^2} \|\mathbf{y}_p - \boldsymbol{g}_b(\mathbf{M}\mathbf{a}_k)\|^2\right] \prod_{r=1}^{R} a_{k,r}^{\eta-1} \qquad (17)$$

where $\mathbf{p}_k = \{p \in \{1, \ldots, P\} | z_p = k\}$. As seen this distribution is too complex to be directly sampled. Here, we use the MCMC sampler to generate samples distributed according to (17). As mentioned earlier, we choose Metropolis-within-Gibbs sampler in which new samples are generated by Gaussian random walk procedure [Roberts 1996].

*Conditional distribution of $c_p$*

Using Bayes theorem, the conditional distribution of class label $c_p$ is expressed by the probabilities

$$\mathbb{P}\left(c_p = k | \mathbf{\theta}_{\backslash c_p}, \mathbf{Y}, \mathbf{M}\right) \propto f\left(\mathbf{y}_p | \mathbf{a}_{c_p}, b, c_p, \sigma^2, \sigma_b^2\right) f(c_p | \mathbf{c}_{\backslash p})$$

$$\propto \exp\left[-\frac{1}{2\sigma^2} \|\mathbf{y}_p - \boldsymbol{g}_b(\mathbf{M}\mathbf{a}_k)\|^2\right] \times \exp\left[\beta \sum_{p' \in \nu(p)} \delta(c_p - c_{p'})\right] \qquad (18)$$

The derived posterior distribution of the class labels of all pixels $\mathbf{c}$ in (18) describes an MRF. Thus, drawing class labels from conditional distribution (18) can be reached by Gibbs sampler in Alg. 1.

*Conditional distribution of $b$*

Using (7) and the Gaussian prior for nonlinear coefficient $b$, the conditional distribution is obtained Gaussian too, as follows

$$f\left(b | \mathbf{\theta}_{\backslash b}, \mathbf{Y}, \mathbf{M}\right) \sim \mathcal{N}(\mu_b, s_b^2) \qquad (19)$$

where

$$\mu_b = \frac{1}{P}\sum_{k=1}^{K}\sum_{p\in \mathbf{p}_k} \frac{\sigma_b^2(\mathbf{y}_p-\mathbf{Ma}_k)^T\mathbf{h}(\mathbf{a}_k)}{\sigma_b^2\mathbf{h}(\mathbf{a}_k)^T\mathbf{h}(\mathbf{a}_k)+\sigma^2} \quad (20)$$

and

$$s_b^2 = \frac{1}{P}\sum_{k=1}^{K}\frac{\sigma_b^2\sigma^2 n_k}{\sigma_b^2\mathbf{h}(\mathbf{a}_k)^T\mathbf{h}(\mathbf{a}_k)+\sigma^2} \quad (21)$$

where $\mathbf{h}(\mathbf{a}_k) = (\mathbf{Ma}_k) \odot (\mathbf{Ma}_k)$, which $\odot$ is Hadamard product, and $n_k$ is number of pixels belong to class label $k$. Consequently, sampling from (19) is done easily.

**Algorithm 1** MRF Implementation

Repeat $N_{MC}$ times:

1. Input: $\mathbf{y}_p, \mathbf{M}, \mathbf{A}, b, c_{p'\in v(p)}, \sigma^2$
2. Output: $c_p$
3. for $k = 1:K$
4.   Compute $w_k \propto \mathbb{P}\left(c_p = k|\boldsymbol{\theta}_{\setminus c_p}, \mathbf{y}_p, \mathbf{M}\right)$ in (15)
5. end
6. Compute the normalizing constant
7.   $G(\beta) = \sum_{k=1}^{K} w_k$
8. Set the probability vector
9.   $\widehat{\mathbf{w}} = \left[\frac{w_1}{G(\beta)}, \ldots, \frac{w_K}{G(\beta)}\right]$
10. Draw $c_p$ in $\{1, \ldots, K\}$ with probability $\{\widehat{w}_1, \ldots, \widehat{w}_K\}$.

*Conditional distribution of $\sigma^2$*

The full-conditional pdf of noise variance is expressed as following based on Bayes theorem:

$$f(\sigma^2|\boldsymbol{\theta}_{\setminus \sigma^2}, \mathbf{Y}, \mathbf{M}) \propto f(\sigma^2)\prod_{p=1}^{P} f\left(\mathbf{y}_p|\mathbf{a}_{c_p}, c_p = k, \sigma^2\right) \quad (22)$$

Due to choosing Jeffrey's prior, the conditional distribution is described as

$$f(\sigma^2|\boldsymbol{\theta}_{\setminus \sigma^2}, \mathbf{Y}, \mathbf{M}) \sim \mathcal{IG}\left(\frac{LP}{2}, \sum_{p=1}^{P}\frac{\|\mathbf{y}_p - g_b(\mathbf{Ma}_{c_p})\|^2}{2}\right) \quad (23)$$

which it is a known distribution that can easily be sampled.

*Conditional distribution of $\sigma_b^2$*

Considering (15), straightforward computations leads one to the following distribution

$$f\left(\sigma_b^2|\boldsymbol{\theta}_{\backslash \sigma_b^2}, \mathbf{Y}, \mathbf{M}\right) \sim \mathcal{IG}\left(\frac{1}{2}+\gamma, \frac{b^2}{2}+\rho\right) \qquad (24)$$

from which it is easy to sample.

To end, using computed conditional distribution of unknown parameters and hyperparameter, a summary of algorithm can be provided as Alg. 2.

**Algorithm 2** Gibbs sampler proposed for hyperspectral unmixing using spatial correlation in PPNMM

1. Input: $\boldsymbol{Y} = [\mathbf{y}_1, \ldots, \mathbf{y}_N], \boldsymbol{M} = [\mathbf{m}_1, \ldots, \mathbf{m}_R], K$
2. Output: $\hat{\mathbf{a}}, \hat{b}, \hat{\mathbf{c}}, \hat{\sigma}^2, \hat{\sigma}_b^2$
3. Initialization:
4. Sample $\mathbf{c}^{(0)}, \mathbf{a}^{(0)}, b^{(0)}, \sigma^{2(0)}, \sigma_b^{2(0)}$ according to their prior distribution
5. Repeat $N_{MC}$ times:
6.    for each pixel $p = 1, \ldots, P$
7.       Sample $\mathbf{a}_{c_p=k}^{(t)}$ according to (17) using MCMC sampler
8.       Sample $\mathbf{c}_p^{(t)}$ according to Alg. 1
9.    end
10.    Sample $b^{(t)}$ according to (19)
11.    Sample $\sigma^{2(t)}$ according to (23)
12.    Sample $\sigma_b^{2(t)}$ according to (24)

Following the explained procedure in Alg. 2, the samples for all unknown parameters would be generated for which an MMSE estimator could be used to compute the sample mean of them as the estimates.

**Experimental Results**

To evaluate the performance of proposed unmixing algorithm, our experiments were performed on both synthetic and real hyperspectral image.

*Simulated Data*

In first evaluation strategy, three $25 \times 25$ pixel hyperspectral images with $K = 3$ classes were generated. These images $I_1$, $I_2$ and $I_3$ were generated by different mixing models including LMM, GBM and PPNMM to evaluate the robustness of algorithms accuracy to different mixing models. The nonlinear parameter $b = 0.1$ was used for PPNMM synthesized image, and similarly the nonlinear parameter $\gamma = [0.5 \quad 0.1 \quad 0.3]$ was established for GBM synthesized image. For each of $K = 3$ label classes, a unique abundance vector was used according to Table 1. An additive Gaussian noise with variance $\sigma^2 = 0.001$ corrupted the synthetic images. This variance corresponds to a reasonable signal to noise ratio around 15 dB in our problem. We select 8 endmembers randomly from the USGS library [Clark 2007] and make our own library. We selected three endmembers to contribute making mixed pixel while five other endmembers were just in library to simulate a semi-supervised scenario. Note that in this case we are not aware of neither the number of endmembers nor the associated ones in the mixing process. The spectral signature of these materials are illustrated in Fig. 2.

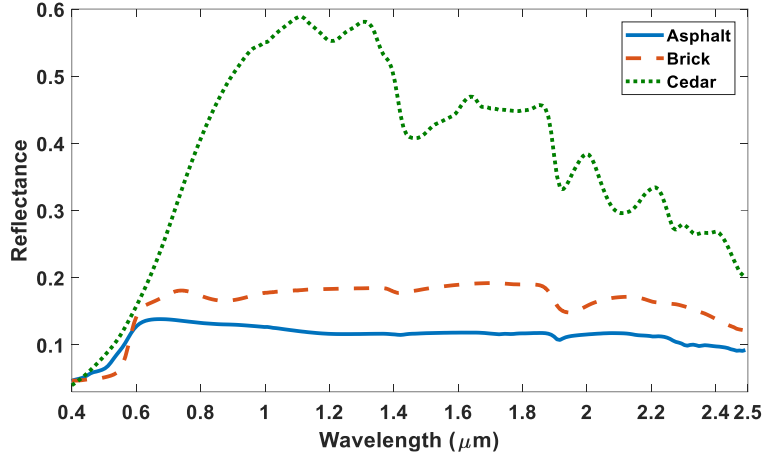

**Figure 2** Spectral signature of three endmembers from USGS [Clark 2007]

A random label map which generated by MRF with $\beta = 1.1$ was assigned to each images. We also choose $\eta = 0.2$ making a sparse distribution. Simulations run with 5000 MCMC and 500 burn-in iterations.

Table 1. Real Abundance Vectors for Three Classes

|         | Real Abundance Vector |
|---------|------------------------|
| Class 1 | $[0.6, 0.1, 0.3, 0, 0, 0, 0, 0]^T$ |
| Class 2 | $[0.1, 0.3, 0.6, 0, 0, 0, 0, 0]^T$ |
| Class 3 | $[0.3, 0.4, 0.3, 0, 0, 0, 0, 0]^T$ |

The real and estimated class labels are illustrated in Fig. 1 (a), 2 and 3 for LMM, GBM and PPNMM synthesized images respectively. The initial random class labels and estimated class map for three first iterations are shown in Fig. 1 (b) for linear mixing model. As seen, the class labels converge to real label map in initial iterations and finally goes to true label map. The results demonstrate that the algorithms estimate the pixel classes perfectly. The classification is done by unmixing simultaneously.

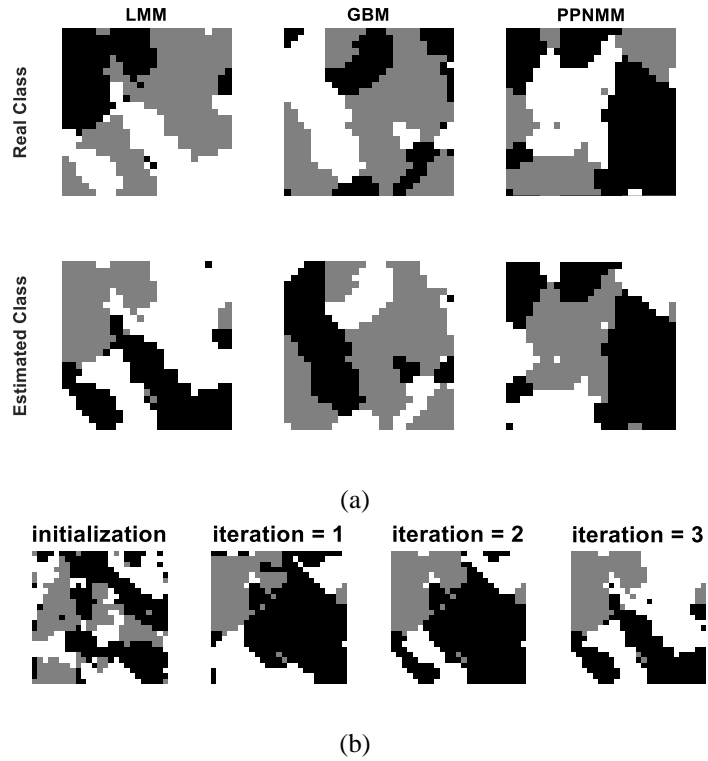

(a)

(b)

Figure 3 Real and estimated label maps (a) for three mixing model synthesized images, and (b) initial and estimated label maps for first three iterations

The estimated abundance values using the proposed MRF-SDP-PPNMM algorithm is presented in Fig. 4*a-c*. As seen, the estimated values are close to the real values. The non-participant endmembers take negligible abundance values in order of $10^{-4}$; which means this endmembers are not exist in hyperspectral pixel.

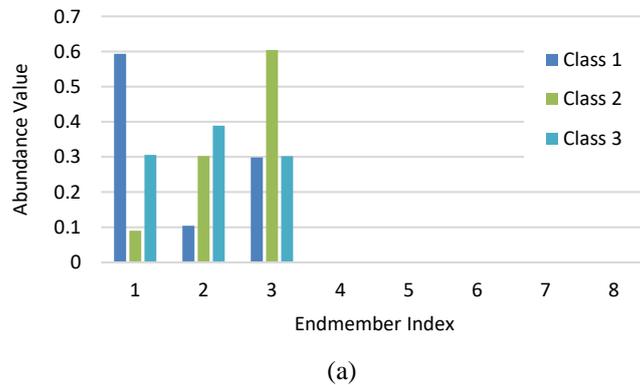

(a)

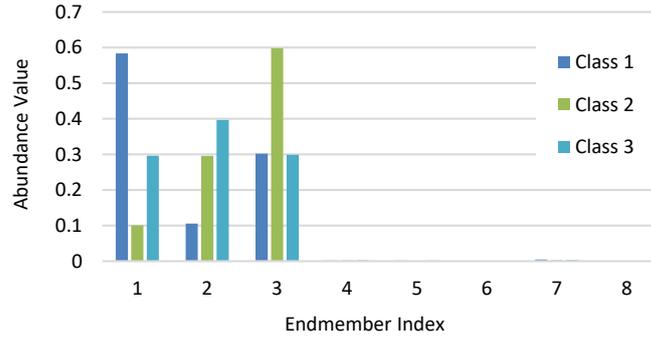

(b)

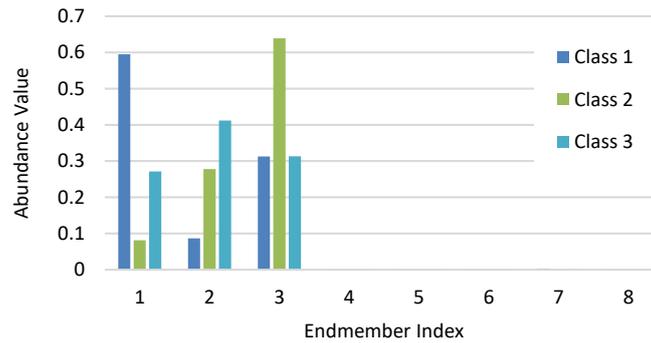

(c)

Figure 4 Estimated abundance vectors of three classes for (a) LMM synthesized image, (b) GBM synthesized image and (c) PPNMM synthesized image

  The Bayesian approaches have the advantage that the full posterior distribution of unknown parameters are achieved during the solution. The pdf of unknown parameter gives us valuable information about the variable. So, the posterior distribution of the nonzero abundance values for three classes are also illustrated in Figs. 3-*a, b*, and *c*. The nonzero abundance values of the first class are 0.6, 0.1 and 0.3 which as seen in Fig. 3*a*, the posteriors are estimated accurately very close to real values. Similarly, the abundances distributions of the second class in Fig. 3b show that the unmixing algorithm could estimate the values 0.1, 0.3 and 0.6 precisely. In third class, as the endmembers were mixed highly, the unmixing algorithm had less accuracy in comparison with two first classes. However, the result are still close to what supposed to be.

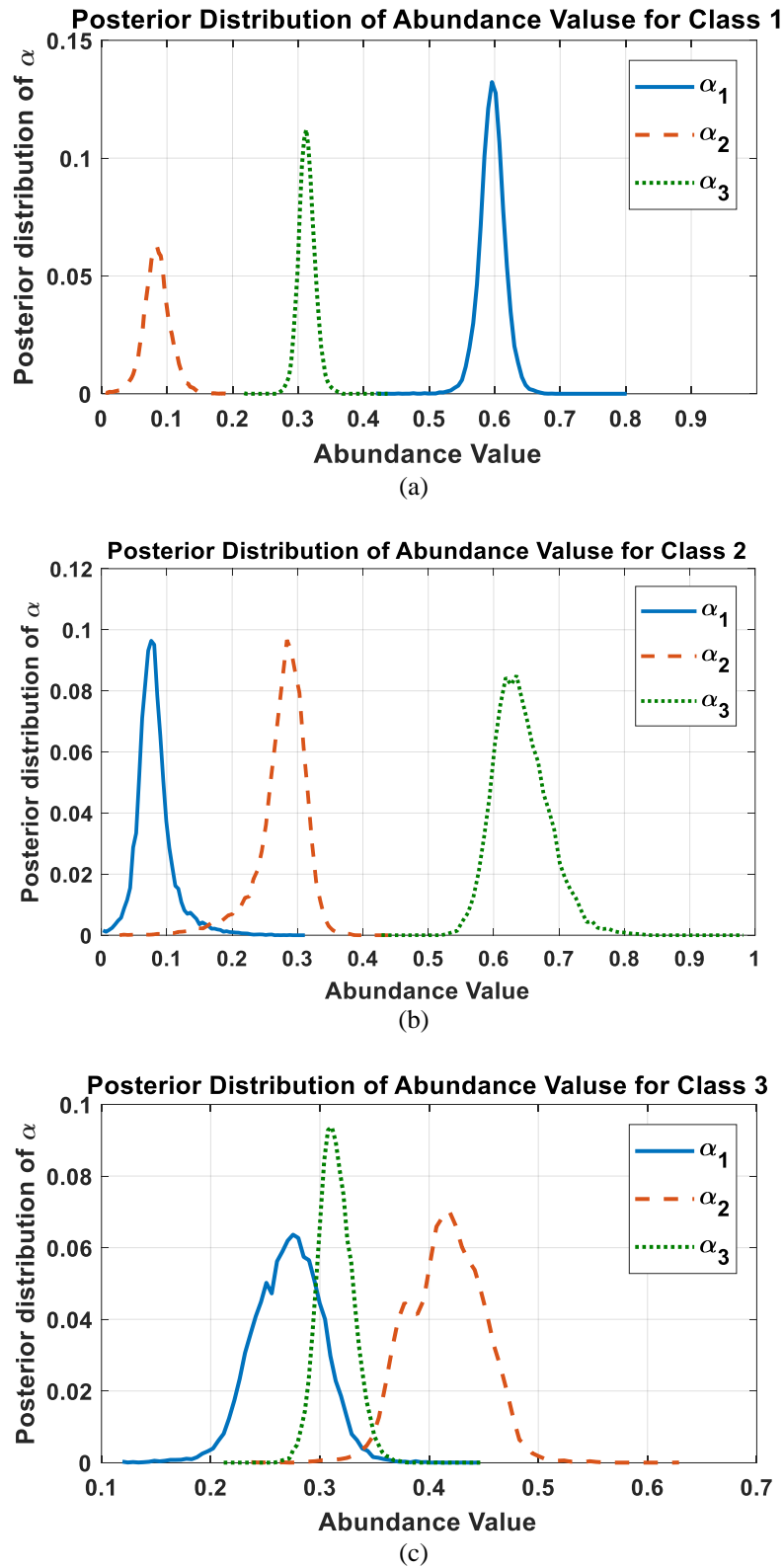

Figure 5 The posterior distribution of abundance values for 3 most significant endmembers for LMM image

To evaluate the unmixing accuracy quantitatively, we also calculate the Root Mean Square Error (RMSE) of abundance estimates and the Reconstruction Error (RE) of the mixed hyperspectral image estimate. The RMSE and RE parameters are defined as follows:

$$\text{RMSE} = \sqrt{\frac{1}{P}\sum_{p=1}^{P}\left\|\hat{\mathbf{a}}_p - \mathbf{a}_p\right\|^2} \qquad (25)$$

$$\text{RE} = \sqrt{\frac{1}{PL}\sum_{p=1}^{P}\left\|\hat{\mathbf{y}}_p - \mathbf{y}_p\right\|^2} \qquad (26)$$

Table 2 shows the RMSE of abundance estimates on three synthesized images for three most common unmixing algorithms. The implemented unmixing algorithms were classified to three categories: the LMM-based solutions, the GBM-based solutions and the PPNMM-based solutions. Since our aim was to have a comprehensive comparison, for each type of mixing model a corresponding appropriate unmixing algorithm was used. We select FCLS unmixing algorithm [Heinz 2001] and the MRF based LMM algorithm [Eches 2011] for LMM. For GBM the Sub-gradient based solution implemented as explained in [Halimi 2011]. Due to getting similar results in Bayesian and Sub-gradient based solutions in [Altmann 2012], we select the faster one for comparison. Also, we use both Bayesian and Sub-gradient based solutions for PPNMM [Altmann 2012]. The results show the preference of the proposed MRF-Bayesian based algorithm.

It's important to note that FCLS, MRF based LMM algorithm, Sub-gradient GBM, Sub-gradient PPNMM and Supervised Bayesian PPNMM algorithm were performed on a supervised manner; which means the exact three endmembers were fed to unmixing procedure. While, in our proposed method a library contains eight endmembers was provided to unmixing process. We evaluate the Bayesian PPNMM algorithm in semi-supervised scenario to only compare the result with our algorithm. The results show that if the materials in library get increased compared to true endmembers, the supervised algorithms would dropped significantly in estimation accuracy. However, our method is not so sensitive to the size of library. As seen in Table 2, best results are acquired by proposed algorithm. Comparison between MRF based unmixing algorithms and the others show that using spatial correlation enhance the unmixing accuracy significantly. Also, using MRF in PPNMM results smaller estimation error compared to the LMM one even in $I_1$ where the synthesized image is made under linear mixing model.

Table 2. Abundance RMSEs on Synthetic Images

|  | $I_1$ (LMM) | $I_2$ (GBM) | $I_3$ (PPNMM) |
|---|---|---|---|
| FCLS [Heinz 2001] | 0.1477 | 0.1471 | 0.1467 |
| LMM (MRF Bayesian) [Eches 2011] | 0.0358 | 0.0567 | 0.0615 |
| GBM (Gradient) [Halimi 2011] | 0.1474 | 0.1465 | 0.1465 |
| PPNMM (Gradient) [Altmann 2012] | 0.2023 | 0.1965 | 0.1973 |
| PPNMM (Bayesian) [Altmann 2012] | 0.1322 | 0.1370 | 0.1270 |
| PPNMM (Bayesian Semi-Sup.) | 0.1995 | 0.2062 | 0.1742 |
| Proposed PPNMM (MRF Bayesian) | **0.0104** | **0.0138** | **0.0315** |

In Table 3 the REs of reconstructed image for three synthesized images were presented. We computed the RE respect to the ideal synthesized image, not to the noisy one. Because if the algorithm had been sensitive to noise, the RE for noisy image would be smaller than the ideal image. As seen, the RE for the MRF-Bayesian is around ten times smaller than other algorithms.

Table 3. Image REs on Synthetic Images

|  | $I_1$ (LMM) | $I_2$ (GBM) | $I_3$ (PPNMM) |
|---|---|---|---|
| FCLS [Heinz 2001] | 0.0028 | 0.0037 | 0.0027 |
| LMM (MRF Bayesian) [Eches 2011] | 0.0021 | 0.0022 | 0.0077 |
| GBM (Gradient) [Halimi 2011] | 0.0028 | 0.0038 | 0.0027 |
| PPNMM (Gradient) [Altmann 2012] | 0.0035 | 0.0043 | 0.0034 |
| PPNMM (Bayesian) [Altmann 2012] | 0.0066 | 0.0076 | 0.0090 |
| PPNMM (Bayesian Semi-Sup.) | 0.0153 | 0.0141 | 0.0141 |
| Proposed PPNMM (MRF Bayesian) | **0.0004** | **0.0013** | **0.0007** |

*Real Data*

The second part of evaluation was performed on real data collected by the Airborne Visible Infrared Imaging Spectrometer (AVIRIS) over Cuprite, Nevada, USA [Vane 1993]. A $50 \times 50$ pixels window was cropped from the whole image and the algorithm was run on it. Both of these images are shown in Fig. After removing the water absorption bands, the number of channels reduced from $L = 224$ to 189 bands.

We select 93 endmembers of the USGS mineral library as our own library, whereas for supervised MRF-LMM algorithm the endmembers are extracted from VCA algorithm [Nascimento 2005] with $R = 14$ as pointed in [Eches 2011]. The algorithm was run with 5000 MCMC iterations and 500 burn-in iterations. We set the number of classes equal to $K = 5$ according to [Eches 2011].

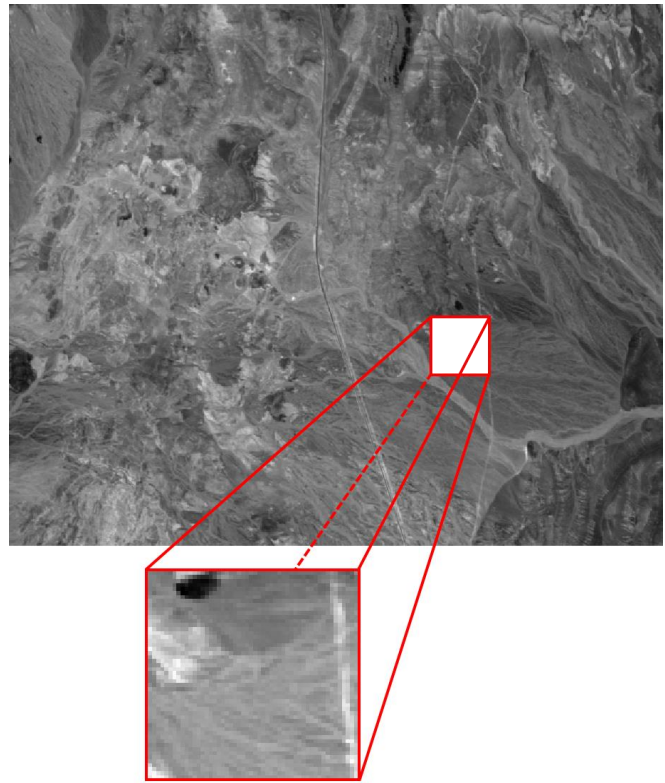

Figure 6 Real hyperspectral images of Cuprite acquired by AVIRIS and the region of interest

The resulted estimated abundance vectors are illustrated in Fig. 6. As seen, most of the abundance values for each class are near zero, so the sparsity constraint is valid. The abundance values for the most significant endmembers are also shown in Table 4 for 5 classes. It's interesting to note that if the number of classes set larger than the actual number of classes, the estimated abundance value corresponding to the same classes would be similar. This result is understanding from Table 4 for the second and fifth classes.

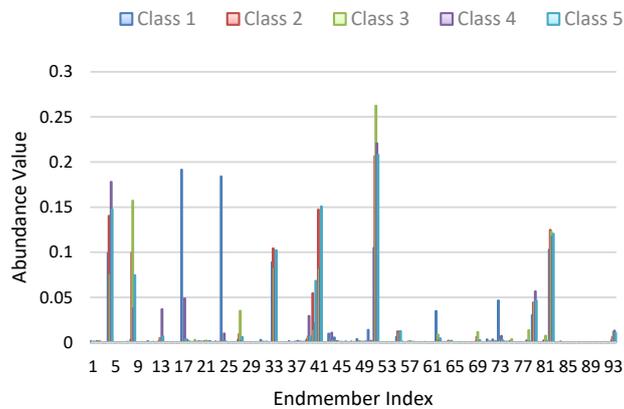

Figure 7 Abundance vectors for 5 classes of Cuprite image

Table 4 Most Significant Abundance Values of 5 Class Labels for Cuprite Image

| Class 1 | $\begin{bmatrix} a_{17} & a_{24} & a_{51} & a_{82} & a_{4} & a_{33} \\ 0.19 & 0.18 & 0.10 & 0.10 & 0.10 & 0.09 \end{bmatrix}$ |
|---|---|
| Class 2 | $\begin{bmatrix} a_{51} & a_{41} & a_{4} & a_{82} & a_{33} & a_{8} \\ 0.21 & 0.15 & 0.14 & 0.12 & 0.10 & 0.10 \end{bmatrix}$ |
| Class 3 | $\begin{bmatrix} a_{51} & a_{8} & a_{82} & a_{33} & a_{41} & a_{4} \\ 0.26 & 0.16 & 0.12 & 0.08 & 0.08 & 0.08 \end{bmatrix}$ |
| Class 4 | $\begin{bmatrix} a_{51} & a_{4} & a_{82} & a_{33} & a_{41} & a_{79} \\ 0.22 & 0.18 & 0.12 & 0.08 & 0.08 & 0.05 \end{bmatrix}$ |
| Class 5 | $\begin{bmatrix} a_{51} & a_{41} & a_{4} & a_{82} & a_{33} & a_{8} \\ 0.21 & 0.15 & 0.15 & 0.12 & 0.10 & 0.07 \end{bmatrix}$ |

We also showed the abundance map for six most significant endmembers in Fig. 8 and the classification map was plotted in Fig. 9. Figure 8 shows that the proportion of materials in disjoint regions are completely different, which this information helps to find the classes of all pixels. Looking at the label maps in Fig. 9, it is obvious that although the proposed MRF-SDP-PPNMM algorithm was used in a semi-supervised scenario against the supervised MRF-LMM algorithm, but it could better assign labels to the image. The bottom half of the image is not classified well by MRF-LMM algorithm, while the MRF-SDP-PPNMM algorithm classify the image appropriately.

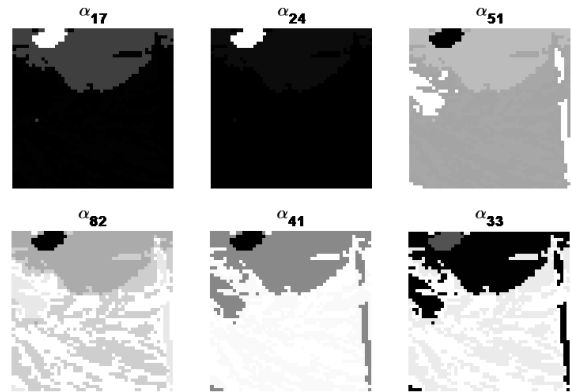

Figure 8 Abundance maps estimated for 6 most significant endmembers form 93 members dictionary on Cuprite image

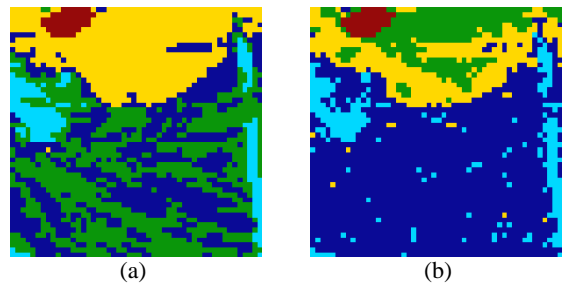

Figure 9 Label maps estimated by (a) proposed MRF-SDP-PPNMM and (b) MRF-LMM algorithm [Eches 2011] with $K = 5$ on Cuprite image

    The reconstruction error is computed for two best algorithms of Table 2 and reported in Table 5. An interesting result was achieved; both algorithms had the same accuracy in the sense of reconstruction error. However there is no measure to compare the abundance estimates because for real images, the exact proportion of materials is not available. It's important to note that although both algorithms result equal error, but the evaluation condition is different; LMM based algorithm is applied in supervised manner, while the proposed algorithm is in semi-supervised scenario. The former one had the

exact 14 endmembers whereas the latter should select 14 endmembers from a library of 93 members. Under this rigid situation the proposed algorithm performed excellently.

Table 5. RE on real Cuprite image

|    | LMM (MRF Bayesian) [Eches 2011] | Proposed PPNMM (MRF Bayesian) |
|----|---------------------------------|-------------------------------|
| RE | 0.0242                          | 0.0242                        |

We repeat the algorithm for a larger hyperspectral image from Cuprite image. The region of interest of size $200 \times 200$ is shown in Fig. 10. We set the number of classes $K = 14$ according to [Eches 2011]. We use the reduced mineral USGS library with 93 endmembers.

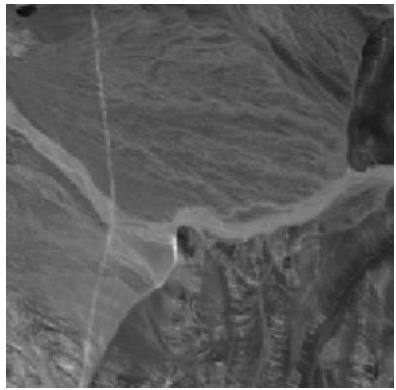

Figure 10 The $200 \times 200$ ROI of Cuprite image

The estimated label map is shown in Fig. 11. As seen the algorithm could labeled the homogenous parts of image accurately.

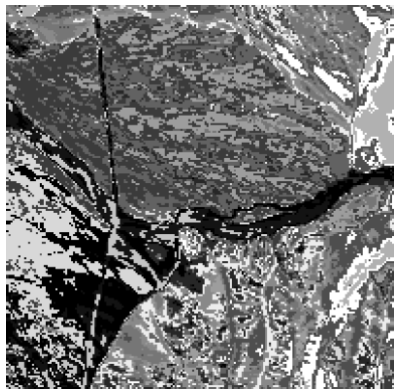

Figure 11 the estimated label map by MRF-SDP-PPNMM with $K = 14$

**Conclusion**

We derived a hierarchical Bayesian algorithm for unmixing of hyperspectral images based on the Polynomial Post-nonlinear Mixing Model. A Dirichlet prior was proposed for modeling the abundance vector sparsity. We set the concentration parameter in such way that the abundance pdf leads to a sparse distribution. In this way, if a huge library is given, the unmixing procedure could be done precisely and any third-party EE algorithm would not be necessary. We also perform Markov Random Fields to the nonlinear mixing model to benefit from spatial correlation to enhance the unmixing process. So hyperspectral classification could be done simultaneous to nonlinear unmixing. Due to complexity of derived posterior distribution, the MCMC method was used to estimate the posterior. Using proposed algorithm, the unmixing accuracy improved compared to LMM, GBM and simple PPNMM algorithms in the sense of RMSE and RE. Also, although we utilize the proposed algorithm in semi-supervised manner, but the nonlinear based MRF results more accurate label map in contrast to linear one.

**References**

Keshava N., Mustard J. F. (2002) - *Spectral unmixing*, IEEE Signal Processing Magazine, vol 29, no. 1, pp. 44-57,. doi: https://doi.org/10.1.1.69.7411.

Somers B., Delalieux S., Stuckens J., Verstraeten W.W., and Coppin P. (2009) - *A weighted linear spectral mixture analysis approach to address endmember variability in agricultural production systems.* Int. J. Remote Sens., vol. 30, pp. 139-147,. doi: https://doi.org/10.1080/01431160802304625.

Settle J. J. and Drake N. A. (1993) - *Linear mixing and the estimation of ground cover proportions*", Int. J. Remote Sens., vol. 14, pp. 1159-1177,. doi: https://doi.org/10.1080/01431169308904402.

Chang C. I. and Heinz D. C. (2000) - *Constrained subpixel target detection for remotely sensed imagery.* IEEE Trans. Geosci. Remote Sens., vol. 38, pp. 1144-1159,. doi: https://doi.org/10.1109/36.843007

Liu J., and Zhang J., (2014) - *Spectral Unmixing via Compressive Sensing*, IEEE Trans. Geosci. Remote Sens., vol. 52, pp. 7099-7110, doi: https://doi.org/10.1109/TGRS.2014.2307573

Vane G., Green R., Chrien T., Enmark H., Hansen E., and Porter W. (1993) - *The airborne visible/infrared imaging spectrometer (AVIRIS).* Remote Sens. Environ, vol. 44, pp. 127-143, doi: https://doi.org/10.1016/S0034-4257(98)00064-9. Availale: https://aviris.jpl.nasa.gov/data/free_data.html.

Heinz D. C. and Chang C.-I. (2001) - *Fully constrained least-squares linear spectral mixture analysis method for material quantification in hyperspectral imagery.* IEEE Trans. Geosci. Remote Sens., vol. 39, no. 3, pp. 529–545, doi: https://doi.org/10.1109/36.911111.

Eches O., Dobigeon N., Mailhes C., and Tourneret J.-Y. (2010) - *Bayesian estimation of linear mixtures using the normal compositional model*. IEEE Trans. Image Process., vol. 19, no. 6, pp. 1403–1413, doi: https://doi.org/10.1109/TIP.2010.2042993.

Fu X., Ma W. K., Bioucas-Dias J. M., and Chan T. H. (2016) - *Semiblind Hyperspectral Unmixing in the Presence of Spectral Library Mismatches*. IEEE Trans. Geosci. Remote Sens., vol. 54, no. 9, pp. 5171-5184, doi: https://doi.org/10.1109/TGRS.2016.2557340.

Akhtar N., Mian A. (2017) - *RCMF: Robust Constrained Matrix Factorization for Hyperspectral Unmixing*. IEEE Trans. Geosci. Remote Sen., vol. 55, no. 6, pp. 3354-3366, doi: https://doi.org/10.1109/TGRS.2017.2669991.

Yu J., Chen D., Lin Y., and Ye S. (2016) - *Comparison of linear and nonlinear spectral unmixing approaches: a case study with multispectral TM imagery*. Int. J. Remote Sens., vol. 38, no. 3, pp. 773-795, doi: https://doi.org/10.1080/01431161.2016.1271475.

Imbiriba T., Bermudez J. C., Richard C. and Tourneret J.-Y. (2016) - *Nonparametric Detection of Nonlinearly Mixed Pixels and Endmember Estimation in Hyperspectral Images*. IEEE Trans. Image Process., vol. 25, no. 3, pp. 1136-1151, doi: https://doi.org/10.1109/TIP.2015.2509258.

Halimi A., Altmann Y., Dobigeon N., and Tourneret J.-Y. (2011) - *Nonlinear unmixing of hyperspectral images using a generalized bilinear model*. IEEE Trans. Geosci. Remote Sens., vol. 49, no. 11, pp. 4153–4162, doi: https://doi.org/10.1016/j.rse.2009.02.003.

Somers B., Cools K., Delalieux S., Stuckens J., Zande D. V., Verstraeten W., and Coppin P. (2009) - *Nonlinear hyperspectral mixture analysis for tree cover estimates in orchards*. Remote Sens. Environ., vol. 113, no. 6, pp. 1183–1193, doi: https://doi.org/10.1016/j.rse.2009.02.003

Hapke B. W. (1981) - *Bidirectional reflectance spectroscopy. I. Theory*. J. Geophys. Res., vol. 86, pp. 3039–3054, doi: https://doi.org/10.1029/JB086iB04p03039.

Liu K.-H., Wong E., and Chang C.-I. (2009) - *Kernel-based linear spectral mixture analysis for hyperspectral image classification.* in Proc. IEEE WHISPERS, Grenoble, France, pp. 1–4, doi: https://doi.org/10.1109/WHISPERS.2009.5289096.

Altmann Y., Dobigeon N., S McLaughlin. and Tourneret J.-Y. (2011) - *Nonlinear unmixing of hyperspectral images using radial basis functions and orthogonal least squares*. in Proc. IEEE IGARSS Conf., pp. 1151–1154, doi: https://doi.org/10.1109/IGARSS.2011.6049401

Altmann Y., Halimi A., Dobigeon N., Tourneret J. (2012) - *Supervised nonlinear spectral unmixing using a postnonlinear mixing model for hyperspectral imagery*. IEEE Trans. Image Process., vol 21, no. 6, pp. 3017–3025, doi: https://doi.org/10.1109/TIP.2012.2187668

Babaie-Zadeh M., Jutten C., Nayebi K. (2001) - *Separating convolutive post non-linear mixtures.* in Proc. 3rd ICA Workshop, San Diego, CA, pp. 138–143, doi: https://doi.org/10.1.1.12.496

Altmann Y., Dobigeon N., Tourneret J.-Y. (2014) - *Unsupervised Post-Nonlinear Unmixing of Hyperspectral Images Using a Hamiltonian Monte Carlo Algorithm.* IEEE Trans. Image Process., vol. 23, no. 6, pp. 2663-2675, doi: https://doi.org/10.1109/TIP.2014.2314022

Heylen R., Burazerovic D., and Scheunders P. (2011) - *Non-linear spectral unmixing by geodesic simplex volume maximization.* IEEE J. Sel. Topics Signal Process., vol. 5, no. 3, pp. 534–542, doi: https://doi.org/10.1109/JSTSP.2010.2088377.

Winter M. (1999) - *Fast autonomous spectral end-member determination in hyperspectral data*. in Proc. 13th Int. Conf. Appl. Geologic Remote Sensing, vol. 2, pp. 337–344, doi: https://doi.org/10.1117/12.366289.

Nascimento J. M. and Bioucas-Dias J. M. (2005) - *Vertex component analysis: A fast algorithm to unmix hyperspectral data*. IEEE Trans. Geosci. Remote Sensing, vol. 43, no. 4, pp. 898–910, doi: https://doi.org/10.1109/TGRS.2005.844293

Chaudhry F., Wu C.-C., Liu W., Chang C.-I., and Plaza A. (2006) - *Pixel purity index-based algorithms for endmember extraction from hyperspectral imagery.* in Recent Advances in Hyperspectral Signal and Image Processing, C.-I Chang, Ed. Trivandrum, Kerala, India: Res. Signpost, , ch. 2.

Dobigeon N., Tourneret J.-Y., and Chang Ch.-I. (2008) - *Semi-Supervised Linear Spectral Unmixing Using a Hierarchical Bayesian Model for Hyperspectral Imagery*. IEEE Trans. Signal Process., vol. 56, no. 7, pp. 2684-2695, doi: https://doi.org/10.1109/TSP.2008.917851.

Plaza A., Martinez P., Perez R., and Plaza J. (2002) - *Spatial/spectral endmember extraction by multidimensional morphological operations*. IEEE Trans. Geosci. Remote Sens., vol. 40, no. 9, pp. 2025–2041, doi: https://doi.org/10.1109/TGRS.2002.802494.

Rogge D., Rivard B., Zhang J., Sanchez A., Harris J., and Feng J. (2007) - *Integration of spatial-spectral information for the improved extraction of endmembers*. Remote Sens. Environ., vol. 110, no. 3, pp. 287–303, doi: https://doi.org/10.1016/j.rse.2007.02.019.

Zhao Ch., Wan X., Zhao G., Cui B., Liu W., and Qi B. (2017) - *Spectral-Spatial Classification of Hyperspectral Imagery Based on Stacked Sparse Autoencoder and Random Forest*, Europ. J. Remote Sens., vol. 50, no. 1, pp. 47-63, doi: https://doi.org/10.1080/22797254.2017.1274566

Eches O., Benediktsson J. A., Dobigeon N., Tourneret J.-Y. (2013) - *Adaptive Markov Random Fields for Joint Unmixing and Segmentation of Hyperspectral Images.* IEEE Trans. Image Process., vol. 22, no. 1, pp. 5-16, doi: https://doi.org/10.1109/TIP.2012.2204270.

Tarabalka Y., Fauvel M., Chanussot J., and Benediktsson J. A. (2010) - *SVM and MRF-based method for accurate classification of hyperspectral images*. IEEE Geosci. Remote Sensing Lett., vol. 7, no. 4, pp. 736–740, doi: https://doi.org/10.1109/LGRS.2010.2047711.

HongLei Y., JunHuan P., BaiRu X. and DingXuan Z. (2013) - *Remote Sensing Classification Using Fuzzy C-means Clustering with Spatial Constraints Based* *on Markov Random Field.* Europ. J. Remote Sens., vol. 46, no. 1. Pp. 305-316, doi: https://doi.org/10.5721/EuJRS20134617.

Eches O., Dobigeon N., Tourneret J.-Y. (2011) - *Enhancing Hyperspectral Image Unmixing With Spatial Correlations.* IEEE Trans. Geosci. Remote Sens., vol. 49, no. 11, pp. 4239-4247, doi: https://doi.org/10.1109/TGRS.2011.2140119.

Chen P., Nelson J. D. B., Tourneret J.-Y. (2017) - *Towards a Sparse Bayesian Markov Random Field Approach to Hyperspectral Unmixing and Classification.* IEEE Trans. Image Process., vol. 26, no. 1, pp. 426-438, doi: https://doi.org/10.1109/TIP.2016.2622401.

Clark R., Swayze G., Wise R., Livo E., Hoefen T., Kokaly R., and Sutley S., (2007). USGS digital spectral library splib06a: U.S. Geological Survey, Digital Data Series 231. [Online]. Available: http://speclab.cr.usgs.gov/spectral.lib06.

Nascimento J. M. P. and Bioucas-Dias J. M. (2009) - *Nonlinear mixture model for hyperspectral unmixing.* in Proc. SPIE Image Signal Process. Remote Sens., vol. 7477, no. 1, p. 747 70I, doi: https://doi.org/10.1117/12.830492

Ng K. W., Tian G. L., and Tang M. L. (2011) - *Dirichlet and Related Distributions: Theory, Methods and Applications.* (Wiley Press, New York)

Robert C. P., and Casella G. (2004) - *Monte Carlo Statistical Methods*," 2nd ed., Springer Verlag, New York.

Roberts G. O. (1996) - *Markov chain concepts related to sampling algorithms.* in Markov Chain Monte Carlo in Practice, W. R. Gilks, S. Richardson, and D. J. Spiegelhalter, Eds. London, U.K.: Chapman & Hall, pp. 259–273,.